\documentclass{article}
\usepackage{amsmath}
\usepackage{graphicx}

\input{tcilatex}

\begin{document}

\title{Figure \textquotedblleft 8\textquotedblright\ gravitational-wave
antenna using a superconducting-core coaxial cable: Continuity equation and
its superluminal consequences}
\date{White paper of November 4, 2010}
\author{R. Y. Chiao, U. C. Merced}
\maketitle

\begin{figure}
\label{antenna} 
\includegraphics[width=6in]{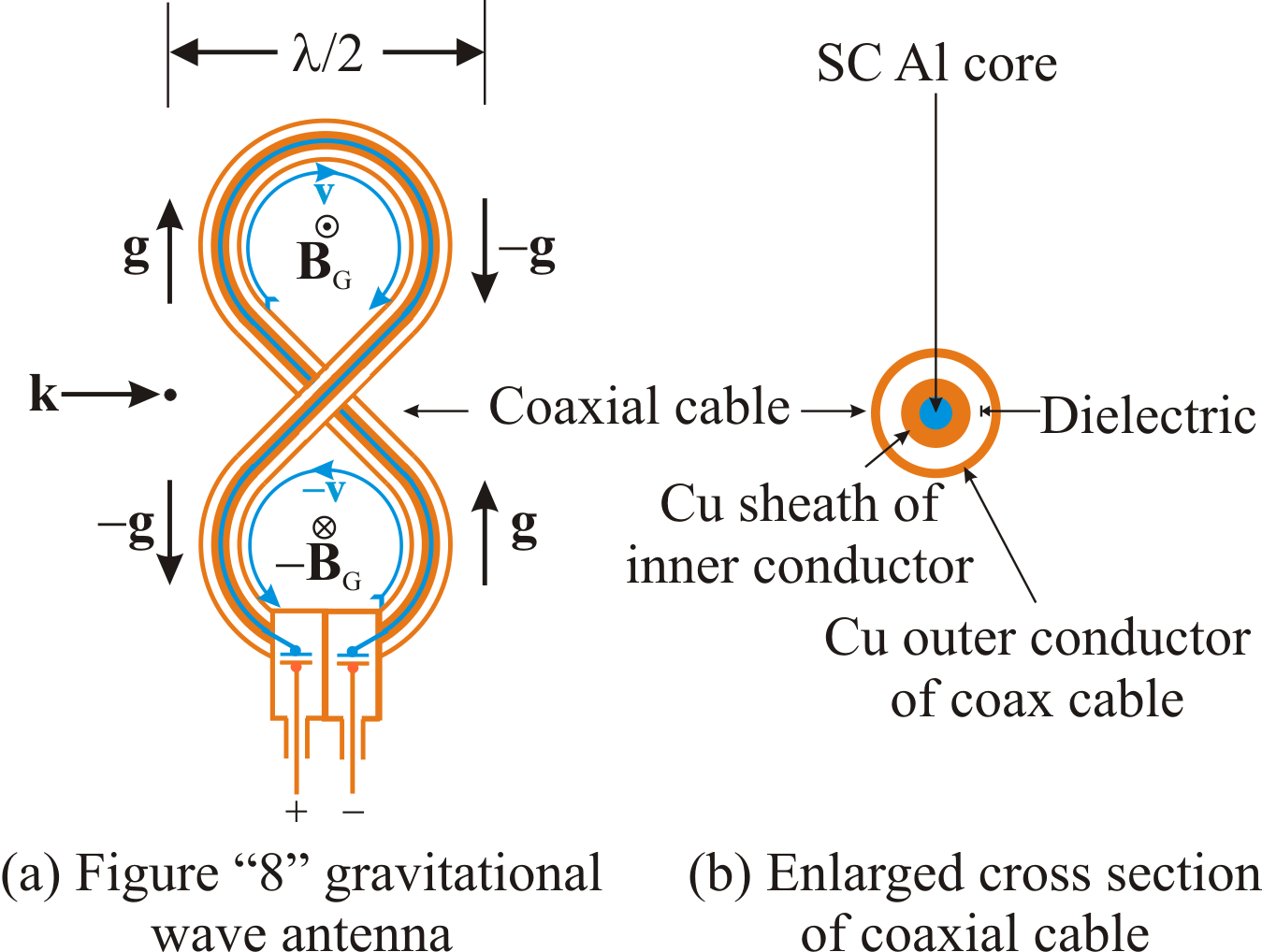}
\caption{(a) Snapshot at $t=0$ of a gravitational plane wave which is
incident with wavevector $\mathbf{k}$ from the left upon a superconducting
(SC)-core coaxial cable that is bent into the shape of a figure
\textquotedblleft 8.\textquotedblright\ The instantaneous local
accelerations at $t=0$ due to gravity are denoted by $\mathbf{g}$ and $%
\mathbf{-g}$ \protect\cite{g=E_G}. The width of the antenna is approximately 
$\protect\lambda /2$, where $\protect\lambda $ is the wavelength of the
gravitational wave. The superfluid velocity of the Cooper pairs is denoted
by $\mathbf{v}$ and the gravito-magnetic field by $\mathbf{B}_{\text{G}}$.
(b) Enlarged cross-sectional view of the coaxial cable. \textit{Blue}
denotes the SC aluminum core of the cable; \textit{orange} denotes the
normal copper sheath surrounding the SC core of the inner conductor, and
also denotes the outer conductor of the cable, as well as the two Faraday
cages which contain the two bimetallic capacitors shown at the bottom of the
figure \textquotedblleft 8\textquotedblright\ in (a).}
\end{figure}

In our Prague paper \cite{Prague}, we predicted that the Cooper pairs will
respond to a gravitational (GR) wave \textit{differently} from the ions of
the ionic lattice of a superconductor (SC), because there exists a global
rigidity of the wavefunction of the former which is absent in the latter.
This results in a \textit{differential} motion between the pairs and the
lattice. Therefore a supercurrent will be induced by the gravitational wave.
This induced supercurrent leads in turn to a charge-accumulation effect at
the boundaries of the SC where the supercurrent terminates. The existence of
the charge-accumulation effect follows from the continuity equation.

Furthermore, we predicted that the supercurrents induced by an incident GR
wave will be \textit{superluminal}. This is one of the most
counter-intuitive results of our Prague paper. We calculated that extremely
fast (i.e., \textquotedblleft instantaneously superluminal\textquotedblright
) electronic mass motions will be induced by a GR wave incident on a square
SC plate. For example, a Cooper pair could disappear from point $A$ at the
edge of the first quadrant of the plate, and then \textit{instantaneously}
re-appear at point $B$ at the edge of the second quadrant of the plate, even
when the points $A$ and $B$ are extremely far away from each other. The
resulting huge mass supercurrents within the SC plate induced by the GR wave
could then lead to a mirror-like reflection of the wave. We shall show here
how such superluminal supercurrents can arise, starting from the continuity
equation.

Consider the SC-core coaxial cable that is bent into the shape of a figure
\textquotedblleft 8\textquotedblright\ shown in Figure 1%
. The normal metallic parts of the cable, indicated in orange (denoting
copper) are transparent to gravitational radiation, which is incident from
the left. As this radiation penetrates into the cable it will encounter its
SC core, indicated in blue (denoting superconducting aluminum). The ions of
the ionic lattice of the cable will undergo inhomogeneous free fall with a 
\textit{local} acceleration due to tidal gravitational forces which are
denoted by $\mathbf{g}$ or $\mathbf{-g}$ in Figure 1,
depending on the location of the ions \cite{Half-wave}.

However, the \textit{non-localizable} Cooper pairs, which have a \textit{%
globally} coherent quantum phase everywhere, will not undergo free fall
along with the \textit{local} motion of the ions. Therefore there will
result a supercurrent which circulates around the figure \textquotedblleft
8\textquotedblright\ in a clockwise sense in the upper half of the figure
\textquotedblleft 8\textquotedblright , but will circulate around in an
anti-clockwise sense in the lower half. This supercurrent will be terminated
by the blue plates of the bimetallic capacitors inside the electronic boxes
are the bottom of the figure \textquotedblleft 8\textquotedblright , and
therefore charges will accumulate at these blue plates. Oppositely signed
charges $+$ and $-$ will then be induced on the two orange plates of these
bimetallic capacitors. The resulting voltage signals can then be fed through
normal SMA\ cables, and detected after amplification using an oscilloscope
or spectrum analyzer.

The conversion efficiency of the incident gravitational wave power into the
electrical signal power depends on the impedance matching conditions of the
figure \textquotedblleft 8\textquotedblright\ antenna. Since the SC has zero
impedance (no losses), when the figure \textquotedblleft
8\textquotedblright\ is short-circuited such that the two blue plates of the
bimetallic capacitors are connected to each other by means of a SC short,
one expects the antenna to behave like a mirror, and reflect or scatter the
incident GR wave efficiently. Hence one expects that it should be possible
to impedance match efficiently the antenna to free space in principle.

If so, by reciprocity, the antenna can serve both as a receiver and as a
generator of GR waves, with an equal efficiency for the generation as for
the reception of GR waves. This implies the possibility of a Hertz-like
experiment, in which two coplanar figure \textquotedblleft
8\textquotedblright\ antennas are placed side by side, their axes being
vertically oriented, with one antenna as the transmitter and the other as
the receiver of microwave-frequency GR waves. This can be done within a
single dilution refrigerator sample can. Since the coaxial cables are
already electrically shielded from each other by the Faraday cages formed by
their outer conductors, any signal communicated from one figure
\textquotedblleft 8\textquotedblright\ to the other can only arise from GR
wave communication. As a control experiment, one can demonstrate that the
signal disappears above the transition temperature.

\begin{figure}
\label{models} %
\includegraphics[width=6in]{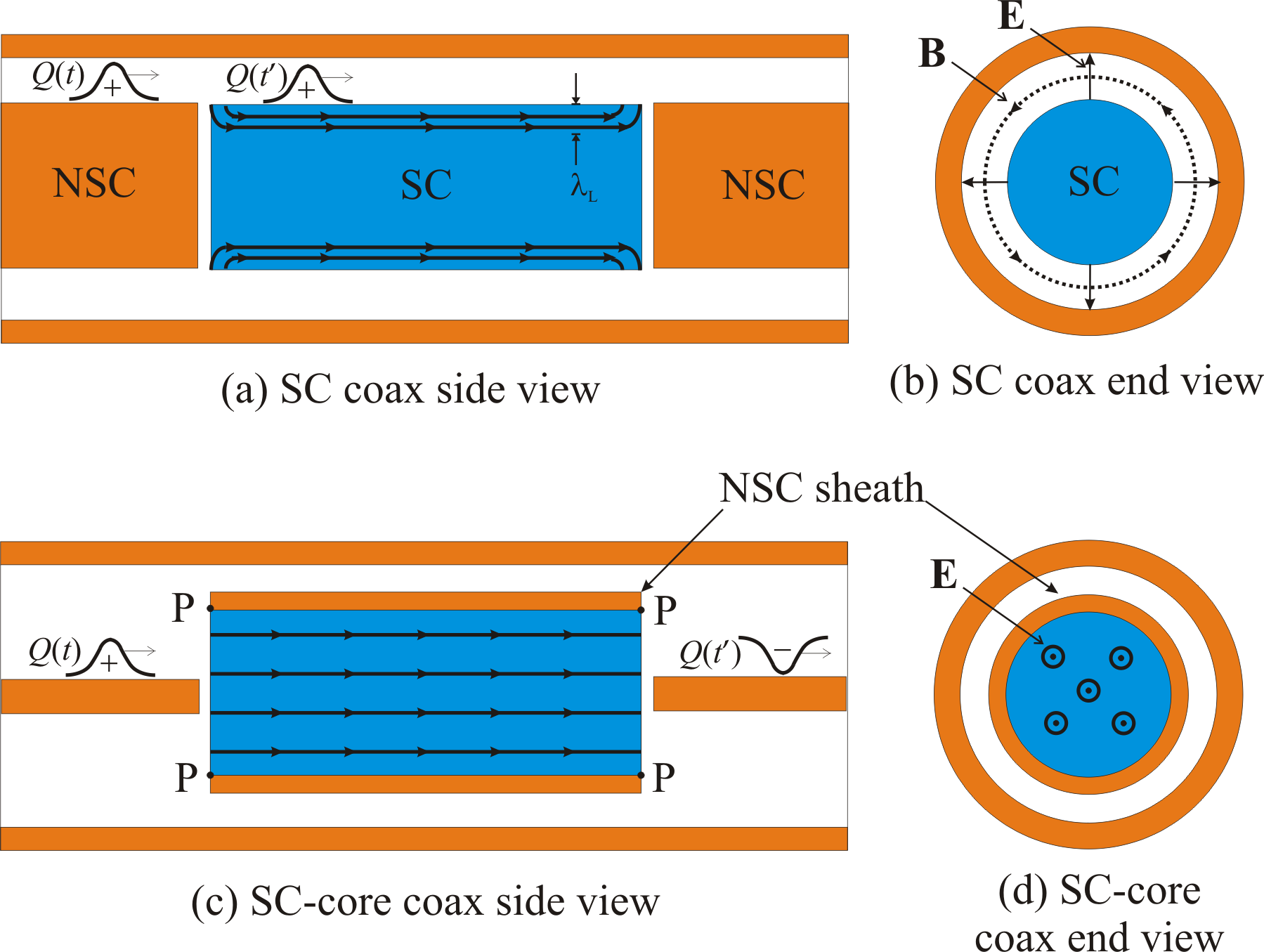}
\caption{A SC coax \textit{without} a NSC sheath is pictured in the upper
half in parts (a) and (b), whereas a SC-core coax \textit{with} a NSC-sheath
is pictured in the lower half of in parts (c) and (d). \textit{Luminal}
pulse propagation of a charge pulse $Q(t)$ occurs in (a) and (b), whereas 
\textit{superluminal} pulse propagation occurs in (c) and (d). In parts (b)
and (d), the electric field lines $\mathbf{E}$ are denoted by solid lines,
and the magnetic field lines $\mathbf{B}$ by dashed lines. In (a), $\protect%
\lambda _{\text{L}}$\ denotes the London penetration depth.}
\end{figure}

In order to understand the operation of the figure \textquotedblleft
8\textquotedblright\ antenna, see Figure 2.
In the upper half of Figure 2, we consider
the usual case of \textit{luminal} propagation at the usual speed of $c/n$
(where $n$ is the refractive index of the dielectric of the cable) of a
charge pulse down a SC coaxial cable with a SC (superconducting) central
section of the cable, but in which the SC section has no NSC
(nonsuperconducting) sheath surrounding it. (\textit{Blue} denotes SC
aluminum; \textit{orange} NSC copper). An incident pulse of charge $Q(t)$ is
incident from the left and crosses the small gaps between the NSC and SC
sections of the cable, because these gaps form high-capacitance capacitors,
which behave like RF shorts for the high frequency pulses like those of $%
Q(t) $ \cite{RF short}. Note that the charges and currents will produce 
\textit{transverse} electric and magnetic fields, which are depicted by the
solid and the dashed lines, respectively, in part (b) of Figure 2. Thus the charge pulse $Q(t)$ will propagate
luminally down the cable towards the right, exciting the usual TEM mode of a
coaxial cable.

In the lower half of this Figure, we consider the unusual case of \textit{%
superluminal} propagation of the charge pulse $Q(t)$ down a SC-core coaxial
cable with the central SC section of the inner conductor intimately
surrounded by a sheath of NSC material. There exists a good electrical
contact between the SC core and the surrounding NSC sheath, so that Cooper
pairs can cross at the point P from the SC side into the NSC side of the
inner conductor, and break up into normal electrons. These electrons will
then diffuse through the NSC sheath to its surface, and travel within a skin
depth of the NSC outer surface of the inner conductor. The thickness of the
NSC sheath will be chosen to be much thicker than the skin depth, so that
the \textit{transverse} electric and magnetic fields of the usual TEM mode
of propagation shown in (b) of the upper half of the Figure, will be shorted
out. Therefore only the \textit{longitudinal} electric fields shown in part
(d) of the lower half of the Figure, will be allowed to propagate down the
cable.

Hence at the point P, there will occur a \textit{partitioning} of charges
into two types: \textit{Type (i)} charges are electrons that result from a
pair-breaking process, in which Cooper pairs cross at P from the SC core
into the NSC sheath, and break up into normal electrons that diffuse towards
the surface of the sheath; \textit{type (ii)} charges are unbroken Cooper
pairs that remain inside the core of the SC cable and propagate along the
equally-spaced straight-line trajectories denoted by the solid black
horizontal lines in part (c) of Figure 2
straight to the other end of the cable. Type (i) charges produce \textit{%
transverse} EM fields, and will therefore travel as a \textit{luminal}
charge pulse down the cable, but type (ii) charges will be shown below to
produce only \textit{longitudinal} electric fields. It will be shown below
that type (ii) charges will travel as a \textit{superluminal} charge pulse
down the cable.

To see why superluminal propagation is possible in lower half of Figure 2, let us start from the continuity equation for
the charged currents associated with the motion of the Cooper pairs, viz.,%
\begin{equation}
\nabla \cdot \mathbf{j}+\frac{\partial \rho }{\partial t}=0
\label{Continuity equation}
\end{equation}%
where $\mathbf{j}$ is the Cooper pairs electrical current density, and $\rho 
$ is their charge density. The physical meaning of this equation is that
charge is conserved during the motion of the Cooper pairs. We define a
time-dependent superflow velocity field $\mathbf{v}$ of the Cooper pairs
through the relationship%
\begin{equation}
\mathbf{j}=\rho \mathbf{v}\text{ ,}
\end{equation}%
where the charge density $\rho $ is related to the complex order parameter $%
\psi $ as follows:%
\begin{equation}
\rho =q\psi ^{\ast }\psi
\end{equation}%
where $q$ is the charge of a Cooper pair.

To avoid the enormous Coulomb energies associated with the unbalanced charge
densities arising from inhomogeneities in the Cooper pair density inside the
SC, one demands that the charge density of the ionic lattice must be \textit{%
exactly} compensated by the charge density of the Cooper pairs at every
point deep inside the bulk of the SC away from the surface \cite{LT}. Since
we will also assume that the ionic lattice possesses a constant, \textit{%
homogeneous} density everywhere inside the SC \cite{Half-wave}, it follows
that the Cooper pair charge density $\rho $ must be a constant of the
motion, viz.,%
\begin{equation}
\rho =q\psi ^{\ast }\psi =\text{ const.}
\end{equation}%
This is consistent with the fact that the ground BCS state of the SC
corresponds to a uniform charge-density state, and the fact that in
first-order perturbation theory, the ground state remains unaltered to
lowest order by perturbations from all radiation fields. It follows that%
\begin{equation}
\frac{\partial \rho }{\partial t}=0\text{ .}
\end{equation}%
Therefore%
\begin{equation}
\rho \nabla \cdot \mathbf{v}+\frac{\partial \rho }{\partial t}=\rho \nabla
\cdot \mathbf{v}=0
\end{equation}%
and therefore that%
\begin{equation}
\nabla \cdot \mathbf{v}=0\,\ .
\end{equation}%
This implies that the superflow velocity field $\mathbf{v}$\ of the Cooper
pairs is \textit{incompressible}.

From DeWitt's minimal coupling rule \cite{Prague}, we showed that%
\begin{equation}
\mathbf{v}=-\frac{q}{m}\mathbf{A}-\mathbf{h}\text{ ,}
\label{superfluid velocity in terms of A and h}
\end{equation}%
where $q$ is the charge and $m$ is the mass of the Cooper pair,
respectively, $\mathbf{A}$ is the vector potential, and $\mathbf{h}$ is
DeWitt's vector potential. The DeWitt (or \textquotedblleft
radiation\textquotedblright ) gauge is being assumed here, with%
\begin{equation}
\nabla \cdot \mathbf{h}=\nabla \cdot \mathbf{A}=0\text{ .}
\end{equation}%
Taking the curl of the superfluid velocity field given by (\ref{superfluid
velocity in terms of A and h}), one obtains%
\begin{equation}
\nabla \times \mathbf{v}=-\frac{q}{m}\nabla \times \mathbf{A}-\nabla \times 
\mathbf{h}=-\frac{q}{m}\mathbf{B}-\mathbf{B}_{\text{G}}\text{ ,}
\end{equation}%
where $\mathbf{B}=\nabla \times \mathbf{A}$ is the magnetic field, and $%
\mathbf{B}_{\text{G}}=\nabla \times \mathbf{h}$ is the gravito-magnetic
field \cite{Prague}. Using Stokes's theorem, one finds that%
\begin{equation}
\doint_{\Gamma }\mathbf{v}\cdot d\mathbf{l}=\dint\limits_{S(\Gamma )}\left(
\nabla \times \mathbf{v}\right) \cdot d\mathbf{S}=-\frac{q}{m}\Phi -\Phi _{%
\text{G}}\text{ ,}
\end{equation}%
where $\Gamma $ is a closed curve chosen to be deep in the middle of the
core of the SC material in the figure \textquotedblleft 8\textquotedblright\
curve of Figure 1, $S(\Gamma )$ is the area enclosed by 
$\Gamma $, $\Phi $ is the time-varying magnetic flux enclosed inside $%
S(\Gamma )$, and $\Phi _{\text{G}}$ is the time-varying gravito-magnetic
flux inside $S(\Gamma )$ induced by the incident GR wave \cite{Flux}. Due to
the presence of the NSC sheath surrounding the SC core of the coax, which
shorts out any high-frequency EM\ fields arising from supercurrents inside
the core, and due to the fact that the gravitational flux $\Phi _{\text{G}}$
is extremely small, it follows that%
\begin{equation}
\doint_{\Gamma }\mathbf{v}\cdot d\mathbf{l}=\dint\limits_{S(\Gamma )}\left(
\nabla \times \mathbf{v}\right) \cdot d\mathbf{S}=0
\end{equation}%
to a very good approximation. For the straight-line sections of SC coax
shown in part (c) of Figure 2, $\mathbf{B}$,
and therefore $\Phi $, also vanish due to the suppression of any
high-frequency EM fields by the NSC sheath surrounding the SC core. Also, $%
\mathbf{B}_{\text{G}}$ and $\Phi _{\text{G}}$ again vanish to a very good
approximation due to the weakness of gravito-magnetic fields.

Therefore the superflow velocity field $\mathbf{v}$ for Cooper pairs inside
the NSC-sheathed SC-core straight section of the coax shown in (c) and (d)
of Figure 2 obeys the two equations%
\begin{eqnarray}
\nabla \times \mathbf{v} &=&0\text{~,}  \label{curl v = 0} \\
\nabla \cdot \mathbf{v} &=&0\text{ .}  \label{div v = 0}
\end{eqnarray}%
In other words, the superflow of Cooper pairs deep inside the middle of the
NSC-sheathed SC core is both \textit{irrotational} and \textit{incompressible%
}. It follows from (\ref{curl v = 0}) that a solution exists of the form%
\begin{equation}
\mathbf{v=\nabla }\varphi
\end{equation}%
for some potential function $\varphi $, and therefore from (\ref{div v = 0})
that%
\begin{equation}
\mathbf{\nabla }^{2}\varphi =0\text{ .}
\end{equation}%
Thus $\varphi $ obeys Laplace's equation, i.e., the Cooper-pair superflow
will be \textit{streamline} flow. In the special case of the laminar
superflow within a straight pipe of constant cross section, the streamlines
are the straight lines indicated in part (c) of Figure 2, which satisfy the 1D solution of Laplace's
equation, viz.,%
\begin{equation}
\varphi (x,y,z,t)=C(t)x\text{ where }C(t)\text{ is constant independent of }%
x,y,z\text{,}  \label{1D solution of Laplace equation}
\end{equation}%
where the $x$ direction has been chosen to coincide with the direction of
the superflow shown in Figure 2, part (c).
The superflow may be \textit{time-dependent} due to the time variations of
the incident charge pulse $Q(t)$; nevertheless, there will exist \textit{%
instantaneous} streamline solutions \textit{everywhere} having the form
given by (\ref{1D solution of Laplace equation}).

We shall call the resulting kind of superluminal effect, \textquotedblleft
instantaneous superluminality\textquotedblright : The charge induced by the
incident GR wave in the figure \textquotedblleft 8\textquotedblright\
antenna squirts out instantaneously from the two ends of the SC-core coax,
as indicated by the $+$ and $-$ signs at the bottom of the figure
\textquotedblleft 8\textquotedblright\ antenna in part (a) of Figure 1.\ Likewise, in order for charge to be conserved, the
polarity of the transmitted charge pulse on the right hand side of the coax
cable in part (c) of Figure 2) must be
reversed in sign with respect to the incident charge pulse.

It is the exponential suppression of the \textit{transverse} EM degrees of
freedom of the SC-core coax cable by the NSC sheath that forces the Cooper
pairs to develop \textit{longitudinal,} Coulomb fields internally within the
SC core in response to the incident charge pulse $Q(t)$ in Figure 2, part (c). This kind of \textquotedblleft
instantaneous superluminality\textquotedblright\ effect associated with
Coulomb fields is highly counter-intuitive. Therefore it is necessary that
we must first check that this effect really exists in the coiled five-meter
SC-core coax cable experiment, but also in the simple thought-experiment to
be presented below.

\section{Appendix: Simple model for demonstrating the possibility of
\textquotedblleft instantaneous superluminality\textquotedblright\ within a
superconducting body}

\begin{figure}
\label{electron-beam-entering-sc-sphere-thru-cu-tube} %
\includegraphics[width=6in]{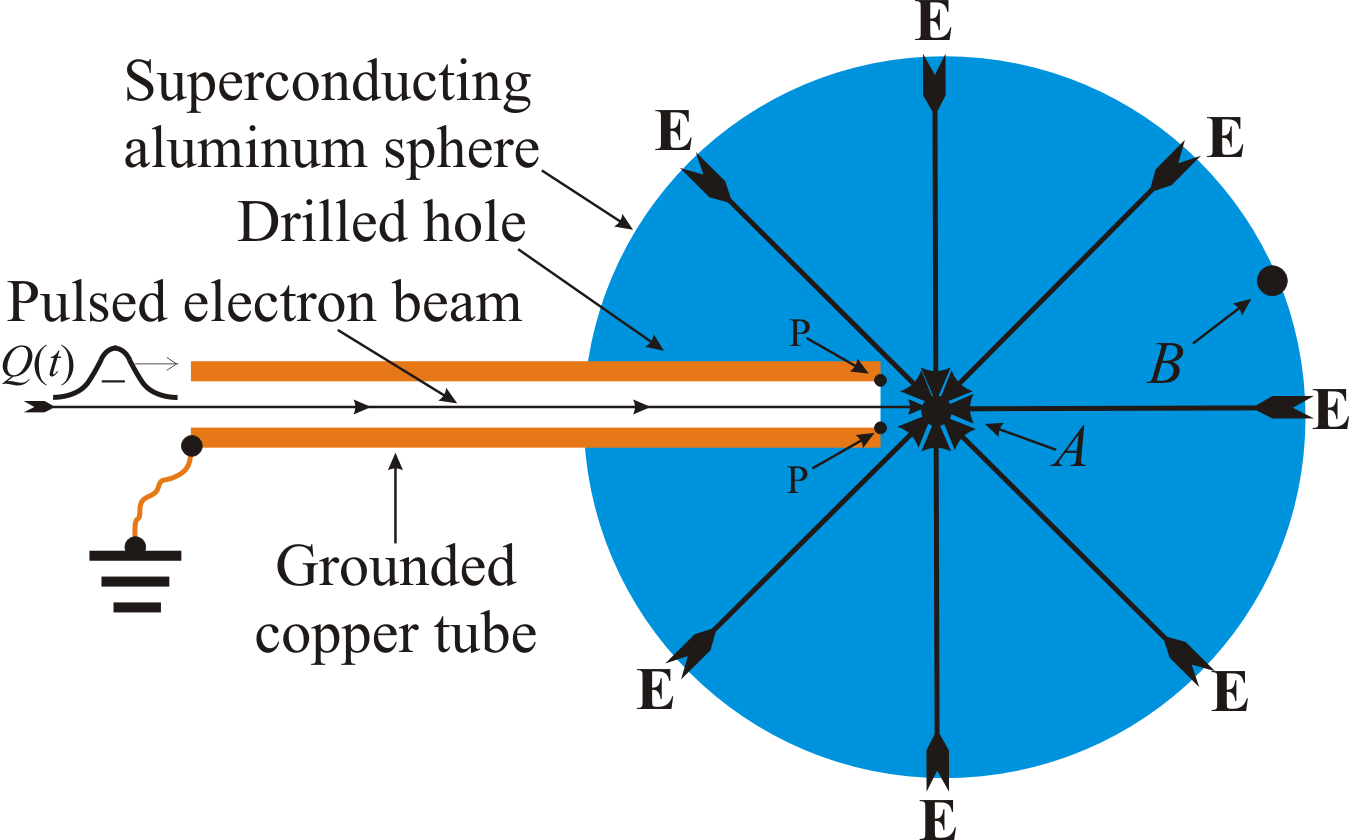}
\caption{A pulsed electron beam enters through a grounded copper tube (in 
\textit{orange}) inserted into a drilled hole that ends just before the
center (point $A$) of a superconducting aluminum sphere (in \textit{blue}).
The beam is stopped by the aluminum, and its charge is deposited at point $A$%
. The deposited charge produces radial lines of electric field $\mathbf{E}$
inside the sphere. How quickly does the deposited charge at point $A$
re-appear at point $B$ as charge on the surface of the sphere? }
\end{figure}

Here we shall show using a simple thought experiment (see Figure 3) how \textquotedblleft
instantaneous superluminality\textquotedblright\ in a SC body follows from
the continuity equation (\ref{Continuity equation}), London's first
equation, and Maxwell's first equation.

In Figure 3, a pulsed
electron beam enters through a long, hollow non-superconducting copper tube,
which is grounded so that it forms a Faraday cage that shields the electrons
in the beam. This tube is inserted into a small hole drilled to the center
of a SC aluminum sphere, in such a way that the copper tube makes an
intimate electrical contact with the aluminum walls of the drilled hole.
Thus the copper tube forms a tightly fitting sleeve inside the hole in the
aluminum sphere. Hence the electrons in the beam cannot \textquotedblleft
see\textquotedblright\ the Cooper pairs in the SC sphere, nor can the Cooper
pairs in the sphere \textquotedblleft see\textquotedblright\ the electrons
in the beam, until the beam strikes the aluminum just in front of the center
of the sphere at point $A$, at which point the electrons come to a sudden
halt. In this way, electrons will be deposited at point $A$.

The Coulomb repulsion between the charges thus deposited at $A$ will drive
them towards the surface of the sphere. However, there are in principle two
different ways by which these charges can reach the surface: They can either
travel as \textit{surface} currents along the inner surface of the copper
tube, or they can travel as \textit{volume} currents within the body of the
aluminum sphere. In the latter case, the charges can be driven towards the
surface by a radial Coulomb field, which are indicated by the radial lines
of electric field $\mathbf{E}$ in Figure 3. The radial $\mathbf{E}$
field can produce supercurrents $\mathbf{j}$ of Cooper pairs within the
volume of the aluminum sphere that can flow out radially towards the surface
of the sphere. In the former case, the Cooper pairs produced by the charge
deposition can flow towards point P of Figure 3\ at the junction between the
NSC copper tube and the SC sphere that is closest to the center at point $A$%
, where they can undergo a pair-breaking process that converts them back
into unpaired electrons that diffuse towards the inner surface of the copper
tube. In this way, the deposited charge from the electron beam at point $A$
can in principle escape as \textit{surface} currents, and thus avoid flowing
as \textit{volume} supercurrents within the interior of the SC body towards
its surface.

At the point P, there will again occur a \textit{partitioning} of charges
into two types: \textit{Type (i)} charges are electrons that result from a
pair-breaking process occurring at P, in which Cooper pairs cross over from
the SC\ aluminum sphere into the NSC copper tube, and will thus break up
into normal electrons that diffuse towards the inner surface of the tube; 
\textit{type (ii)} charges are unbroken Cooper pairs that remain inside the
volume of the SC sphere, and propagate along radial trajectories towards the
surface of the sphere. Again, type (i) charges can in principle produce 
\textit{transverse} EM fields, and will therefore travel as a \textit{luminal%
} charge pulse propagating within the copper tube, whereas type (ii) charges
will be shown below to produce only \textit{longitudinal} electric fields
associated with a \textit{superluminal} charge pulse propagating within the
aluminum sphere.

However, if the inner diameter of the copper tube is chosen to be small,
there will exist a high cutoff frequency of the fundamental TM$_{01}$ mode
of propagation of EM waves within the tube. If the spectrum of frequencies
in the incident Gaussian charge pulse lies well below this cutoff frequency,
the luminal propagation process by type (i) charges along the inner surface
of the copper tube will be exponentially suppressed in favor of the
superluminal propagation process by type (ii) charges within the volume of
the aluminum sphere. The electrons in the incoming electron beam, which can
have a very short wavelength, are not affected by this waveguide-type cutoff
for EM waves.

Furthermore, charges originating from point $A$ can only reach the ground at
the far left end of the copper tube by traveling as luminal type (i) charges
along the outer surface of the section of the copper tube that sticks out
from the surface of the aluminum sphere in Figure 3. Thus there could exist a
considerable dwell time on the surface of the aluminum sphere for the
superluminal type (ii) charges, which could appear instantaneously on the
surface of the sphere, but before these charges could leak into the ground,
they would have to travel as luminal type (i) charges along the outer
surface of an \textit{arbitarily long}, exposed section of the copper tube,
which extends all the way from the surface of the sphere to the tube's far
left end, where the connection to ground is made.

To start the analysis of the propagation of the type (ii) charges, let us
assume that London's first equation holds inside the volume of the SC
aluminum sphere, so that everywhere inside the sphere%
\begin{equation}
\mathbf{j}=-\Lambda \mathbf{A}  \label{London's 1st equation}
\end{equation}%
where $\Lambda $ is London's constant, and $\mathbf{A}$\ is the vector
potential inside the body, which obeys the London (or radiation) gauge%
\begin{equation}
\nabla \cdot \mathbf{A}=0\text{ .}  \label{div A = 0}
\end{equation}%
Therefore the electric field inside the SC is related to the vector
potential by%
\begin{equation}
\mathbf{E}=-\frac{\partial \mathbf{A}}{\partial t}\text{ .}
\label{E related to A}
\end{equation}%
Taking the time derivative of the continuity equation (\ref{Continuity
equation}), one obtains%
\begin{equation}
\nabla \cdot \left( \frac{\partial \mathbf{j}}{\partial t}\right) +\frac{%
\partial ^{2}\rho }{\partial t^{2}}=\nabla \cdot \left( -\Lambda \frac{%
\partial \mathbf{A}}{\partial t}\right) +\frac{\partial ^{2}\rho }{\partial
t^{2}}=\nabla \cdot \left( \Lambda \mathbf{E}\right) +\frac{\partial
^{2}\rho }{\partial t^{2}}=0\text{ .}
\end{equation}%
Assuming that the superconductor is a homogeneous body, so that $\Lambda $
is independent of position inside the SC, one concludes that%
\begin{equation}
\nabla \cdot \left( \Lambda \mathbf{E}\right) +\frac{\partial ^{2}\rho }{%
\partial t^{2}}=\Lambda \nabla \cdot \mathbf{E}+\frac{\partial ^{2}\rho }{%
\partial t^{2}}=\frac{\Lambda }{\varepsilon _{0}}\rho +\frac{\partial
^{2}\rho }{\partial t^{2}}=0\text{ ,}
\end{equation}%
where we have used Maxwell's first equation $\nabla \cdot \mathbf{E}=\rho
/\varepsilon _{0}$. There results a second-order partial differential
equation in time%
\begin{equation}
\frac{\partial ^{2}\rho }{\partial t^{2}}+\frac{\Lambda }{\varepsilon _{0}}%
\rho =0
\end{equation}%
which can be rewritten in the form of a simple harmonic oscillator equation
of motion%
\begin{equation}
\frac{\partial ^{2}\rho }{\partial t^{2}}+\omega _{p}^{2}\rho =0
\label{2nd order PDE}
\end{equation}%
for some natural frequency $\omega _{p}$ which is given by%
\begin{equation}
\omega _{p}=\left( \frac{\Lambda }{\varepsilon _{0}}\right) ^{1/2}\text{ .}
\end{equation}%
The general solution to the PDE (\ref{2nd order PDE}) is%
\begin{equation}
\rho (\mathbf{r},t)=A(\mathbf{r})\cos \omega _{p}t+B(\mathbf{r})\sin \omega
_{p}t  \label{general solution for rho}
\end{equation}%
where the coefficients $A(\mathbf{r})$ and $B(\mathbf{r})$ are to be
determined by the initial conditions inside the SC sphere. Taking the first
partial derivative with respect to time of (\ref{general solution for rho}),
one finds that%
\begin{equation}
\dot{\rho}(\mathbf{r},t)=-\omega _{p}A(\mathbf{r})\sin \omega _{p}t+\omega
_{p}B(\mathbf{r})\cos \omega _{p}t\text{ .}  \label{rho dot}
\end{equation}%
Letting the time $t$ approach the distant past by taking the asymptotic
limit $t\rightarrow -\infty $ in (\ref{general solution for rho})\ and (\ref%
{rho dot}), one determines that at an arbitrary time $t$ in the distant past
and at every point $\mathbf{r}$ in the interior of the SC sphere 
\begin{eqnarray}
\lim_{t\rightarrow -\infty }\rho (\mathbf{r},t) &=&A(\mathbf{r})\cos \omega
_{p}t+B(\mathbf{r})\sin \omega _{p}t=0\text{ ;} \\
\lim_{t\rightarrow -\infty }\dot{\rho}(\mathbf{r},t) &=&-\omega _{p}A(%
\mathbf{r})\sin \omega _{p}t+\omega _{p}B(\mathbf{r})\cos \omega _{p}t=0.
\end{eqnarray}%
Solving for the coefficients $A(\mathbf{r})$ and $B(\mathbf{r})$, one finds
that 
\begin{equation}
A(\mathbf{r})=0\text{ for all }\mathbf{r}\text{ ;}
\end{equation}%
\begin{equation}
B(\mathbf{r})=0\text{ for all }\mathbf{r}\text{ .}
\end{equation}

This is because in the distant past, as one takes the limit $t\rightarrow
-\infty $, long before the arrival of the main peak of the charge pulse ---
which we shall assume for simplicity to be a Gaussian pulse whose peak
arrives at $t=0$ --- the SC body is electrically neutral everywhere, so that
one requires that $\rho =0$ as $t\rightarrow -\infty $ for all $\mathbf{r}$,
and also because in the distant past in the same limit $t\rightarrow -\infty 
$, the body's charge \textit{remains} zero everywhere (i.e., it \textit{stays%
} electrically neutral in the distant past everywhere), so that one also
requires that the first derivative of $\rho $ with respect to time will also
zero as $t\rightarrow -\infty $ for all $\mathbf{r}$. This implies at all
later times and for all points $\mathbf{r}$, except for points at the center
of the sphere, where the charge of electron beam is being deposited, and
except for the points on the surface of the sphere, where the deposited
charge can reappear, that%
\begin{equation}
\rho (\mathbf{r},t)=A(\mathbf{r})\cos \omega _{p}t+B(\mathbf{r})\sin \omega
_{p}t=0\text{ for all }\mathbf{r}\text{\ for all times }t>-\infty
\end{equation}%
including when the main peak of the Gaussian pulse arrives at $t=0$. This
implies that no charge can accumulate at any point $\mathbf{r}$ in the
interior of the SC sphere at any time $t$. Physically, this arises from the
fact that the lines of superflow of the supercurrent $\mathbf{j}$, which
form streamlines, cannot terminate at any point $\mathbf{r}$\ within the
interior of the SC body. Therefore no charge can accumulate at any interior
point $\mathbf{r}$ at any time $t$, and the charge density $\rho $ at $%
\mathbf{r}$ will remain zero for all $t$. In other words, the charge of the
ionic lattice will always be \textit{exactly} compensated by the charge of
the Cooper pairs at all interior points within the body.

Hence the only places in the SC body where charge can accumulate, and
therefore where the charge density can change with time, is either at the
center of the sphere, where the charge from the incoming Gaussian pulse of
electrons is being deposited, or at points on the surface of the sphere
where this deposited charge re-appears in just such way that the \textit{%
total} charge of the entire system is always \textit{exactly} conserved.
This implies that the disappearance of a given electron at point $A\,$is
always accompanied by its \textit{simultaneous} reappearance at an
arbitrarily far-away point $B$ on the surface at \textit{exactly} the same
instant of time. Otherwise, the principle of charge conservation would be
violated.

We shall call this counter-intuitive effect \textquotedblleft instantaneous
superluminality within a SC body.\textquotedblright\ Note that this
superluminal effect does not violate relativistic causality because the
incident charge pulse has an analytic waveform, e.g., a Gaussian, with a
finite bandwidth (i.e., with frequencies less than the BCS gap). There
exists no discontinuous \textquotedblleft front\textquotedblright\ within
the Gaussian waveform, before which the waveform is \textit{exactly} zero.
Such a \textquotedblleft front\textquotedblright\ would contain infinitely
high frequency components that would exceed the BCS gap frequency, and thus
destroy the superconductivity of the sphere. Thus this \textquotedblleft
instantaneously superluminal\textquotedblright\ effect has similarities with
that of a Gaussian wavepacket tunneling through a tunnel barrier in quantum
mechanics, whose early analytic tail contains all the information needed to
reconstruct the entire transmitted wave packet, including its peak, earlier
in time \textit{before} the incident peak could have arrived at a detector
traveling at the speed of light \cite{Chiao-Steinberg}.

What is the London constant $\Lambda $? Recall that DeWitt's minimal
coupling rule (\ref{superfluid velocity in terms of A and h}), when one sets 
$\mathbf{h=0}$ everywhere, leads to the following expression for the
superfluid velocity:%
\begin{equation}
\mathbf{v}=-\frac{q}{m}\mathbf{A}\text{ .}  \label{v related to A}
\end{equation}%
The supercurrent density $\mathbf{j}$ is then related to $\mathbf{v}$ by%
\begin{equation}
\mathbf{j}=\rho \mathbf{v}=nq\mathbf{v=}-\frac{nq^{2}}{m}\mathbf{A}\text{ }%
=-\Lambda \text{ }\mathbf{A}\text{,}
\end{equation}%
where $n$ is the number density of Cooper pairs in the SC and $\rho =nq$ is
their charge density. It follows that London's constant is given by%
\begin{equation}
\Lambda =\frac{nq^{2}}{m}
\end{equation}%
and therefore that the natural frequency%
\begin{equation}
\omega _{p}=\left( \frac{\Lambda }{\varepsilon _{0}}\right) ^{1/2}\text{ }%
=\left( \frac{nq^{2}}{m\varepsilon _{0}}\right) ^{1/2}\text{ }
\end{equation}%
is simply the SC plasma frequency. However, the plasma frequency $\omega
_{p} $\ lies in the UV part of the EM\ spectrum, which is far higher in
frequency than the BCS gap frequency. It should be kept in mind that the
process of pair-breaking at such high UV frequencies would prevent any 
\textit{real} plasma excitations within the SC body from occurring at $%
\omega _{p}$. Nevertheless, \textit{virtual} excitations of the SC plasma at
frequencies lower than the BCS gap frequency, for example, within the RF
part of the EM spectrum, can still occur inside the SC, provided that the
spectrum of the charge pulse incident on the SC sphere in Figure 3 has an upper frequency
cutoff which lies well below the BCS gap frequency.

In order to understand the concept of \textquotedblleft virtual plasma
excitation,\textquotedblright\ let us consider an analogy of (\ref{2nd order
PDE}) with the SHO equation of motion%
\begin{equation}
\frac{d^{2}x}{dt^{2}}+\omega _{0}^{2}x=f(t)
\end{equation}%
where $x$ is the displacement of the oscillator, $\omega _{0}$ is its
resonance frequency, and $f(t)$ is a forcing function whose spectrum of
frequencies lies well below $\omega _{0}$.\ (We assume an oscillator with a
unit mass here.) Then the displacement of the oscillator at the low
frequencies of the driving force $f(t)$ will \textquotedblleft adiabatically
follow\textquotedblright\ this force at each instant of time $t$, so that to
a good approximation, the low-frequency solution is given by%
\begin{equation}
x(t)\approx \frac{f(t)}{\omega _{0}^{2}}\text{ .}
\end{equation}%
There is neither a time-delay nor an energy loss in the system's response to
the driving force, i.e., there is no phase lag nor dissipation in the driven
SHO. This can also happen here with the RF-frequency Gaussian charge pulse.

The term \textquotedblleft virtual\textquotedblright\ is being used in a
quantum sense, in which a non-dissipative \textquotedblleft
virtual\textquotedblright\ transition occurs within a two-level atomic
medium when it is driven by an EM wave whose frequency is tuned far below
the resonance frequency $\omega _{0}$ of the atoms. However, these
\textquotedblleft virtual\textquotedblright\ transitions, which are driven
by the low-frequency EM wave, can still lead to a nontrivial dielectric
constant of the medium. \textquotedblleft Virtual\textquotedblright\
transitions are to be distinguished from \textquotedblleft
real\textquotedblright\ transitions in a system of two-level atoms, in which 
\textit{dissipation} of the EM wave occurs during the \textit{absorption} of
on-resonance photons by the atoms. Likewise, here \textquotedblleft virtual
plasma excitations\textquotedblright\ can lead to \textit{non-dissipative}
charge-accumulation effects (or equivalently, charge polarization effects)
induced on the SC body. No \textit{real} plasma excitations occur within the
SC, but instantaneous Cooper-pair relocations (i.e., \textquotedblleft
instantaneous superluminality\textquotedblright ) can occur through these 
\textit{virtual} interactions.

In connection with the distinction between superluminal and luminal effects,
it is important to introduce the conceptual distinction between \textit{%
longitudinal} and \textit{transverse} supercurrents. Let us define two types
of supercurrent as follows:%
\begin{eqnarray}
\mathbf{j}_{\parallel }(\mathbf{r},t) &\parallel &\mathbf{E}(\mathbf{r},t)%
\text{ as \textquotedblleft longitudinal supercurrents\textquotedblright ;}
\\
\mathbf{j}_{\perp }(\mathbf{r},t) &\perp &\mathbf{E}(\mathbf{r},t)\text{ as
\textquotedblleft transverse supercurrents\textquotedblright .}
\end{eqnarray}%
For example, longitudinal supercurrents can be produced by the time-varying
Coulomb field emanating from the charge being deposited by the pulsed
electron beam at the center of the SC sphere, as illustrated in Figure 3. Such time-varying
longitudinal supercurrents generated by the Coulomb field do\ not
necessarily produce radiation, and therefore one need not expect the usual
retardation effects due to the finite speed of light, which are usually
associated with the radiation produced by transverse supercurrents. In fact,
it is not true that all time-varying supercurrents must lead in all cases to 
\textit{retarded} interactions. As an explicit counter-example, one can
point to the longitudinal supercurrents that give rise to the above
\textquotedblleft instantaneous superluminality\textquotedblright\ effect.

By contrast, transverse supercurrents will be produced by time-varying EM
radiation fields, such as those in the fundamental TEM mode of propagation
down a SC coaxial cable, in which the SC inner conductor does not possess
any NSC (non-superconducting) sheath surrounding it. Transverse
supercurrents will propagate within a London penetration depth $\lambda _{%
\text{L}}$\ of the outer surface of the SC inner conductor, and these
supercurrents are perpendicular to the local electric fields which are
produced through the action of the time-varying magnetic fields generated by
these supercurrents, as illustrated in parts (a) and (b) of Figure 2. Hence time-varying transverse supercurrents
can produce EM radiation fields, as well as being produced by them.
Therefore, in general, one should expect that the retardation effects that
are retarded by the finite speed of light will occur in connection with
these transverse supercurrents.

To sum up, \textit{longitudinal} supercurrents can give rise to \textit{%
superluminal} effects, but \textit{transverse} supercurrents can give rise
to \textit{luminal} effects.


\begin{thebibliography}{9}
\bibitem{g=E_G} The instantaneous local acceleration due to gravity $\mathbf{%
g}$ here is identical to the gravito-electric field $\mathbf{E}_{\text{G}}$%
introduced in \cite{Prague} in connection with the Maxwell-like equations
for gravitational fields.
\bibitem{RF short} The two capacitors (i.e., the two small gaps in part (a)
of Figure 2) behave like shorts for the
RF-frequency pulse $Q(t)$, and will therefore transmit the RF pulse from the
NSC to the SC sections of the cable through the left gap, and then from the
SC to the NSC sections of the cable through the right gap. The RF current
within a skin depth of the surface of the inner conductor of the left NSC
section\ will continue across the left gap from the NSC section into the SC
section of the cable, due to Maxwell's displacement current, and then flow
within a London penetration depth of the surface of the inner conductor of
the SC section. Thus the transmitted pulse $Q(t^{\prime })$ in part (a) of
Figure 2 will continue to propagate to the
right. The initial neutrality of the SC section of the cable is preserved
after the charge pulse $Q(t^{\prime })$ leaves the left gap, by the
development of an exactly compensating oppositely signed charge inside the
right plate of the left capacitor. This compensating charge, which arises
from Maxwell's displacement current, will remain there. The incident charge
pulse $Q(t)$ arriving from the left will deposit all of its charge into the
left plate of the left capacitor, and also remain there. Similarly when the
transmitted charge pulse $Q(t^{\prime })$ arrives at the right gap, it will
deposit all of its charge into the left plate of the right capacitor, and
remain there. An exactly compensating oppositely signed charge will develop
inside the right plate of the right capacitor, which again arises from
Maxwell's displacement current, and a transmitted charge pulse $Q(t^{\prime
\prime })$ will then leave the right capacitor and continue to propagate to
the right within a skin depth of surface of the inner conductor. Thus the SC
coax cable will behave not much differently from a normal coax cable with a
single, continuous NSC inner conductor. If the two gaps are very small, the
energy required to charge the two capacitors up, which must be taken out
from the transmitted pulses, will also be very small, so that there will be
very little diminution of the transmitted pulse energy through the SC coax
cable.
\bibitem{LT} By points \textquotedblleft deep inside\textquotedblright\ the
SC, we mean points many coherence lengths $\xi _{0}$ away from the surface
of the SC. Also, we assume that we are operating at sufficiently low
temperatures so that there are essentially no normal electrons left inside
the SC, because they have all condensed into Cooper pairs.
\bibitem{Flux} The gravito-magnetic flux $\Phi _{\text{G}}$\ subtended by
the figure \textquotedblleft 8\textquotedblright\ antenna will in general be
nonzero, since the contributions to the flux contained inside the upper and
lower loops of antenna add, and do not cancel, due to symmetry (see Figure 1). The gravito-magnetic field $\mathbf{B}_{\text{G}}$\
is maximum in the middle of these loops because a standing GR wave will in
general be set up by reflections of the incident traveling wave from the
left and right sides of the figure \textquotedblleft 8\textquotedblright ,
since they will be separated by half a GR wavelength.
\end{thebibliography}
\end{document}